\documentclass[a4paper,10pt]{article}

\usepackage[T1]{fontenc}

\pdfoutput=1 

\makeatletter
\def\fnum@figure{\textbf{\figurename\nobreakspace\thefigure}}
\def\fnum@table{\textbf{\tablename\nobreakspace\thetable}}
\makeatother

\RequirePackage{doi}

\usepackage{xcolor}
\usepackage{hyperref}

\makeatletter
\def\p@figure{\color{blue}}
\def\p@table{\color{blue}}
\def\p@equation{\itshape\color{violet}}
\makeatother

\usepackage{amsmath}
\usepackage{authblk}
\usepackage{graphicx}
\usepackage{wrapfig}
\usepackage{color, colortbl}
\usepackage{makecell}
\usepackage{multirow}
\usepackage{verbatim}
\usepackage{url}

\newcommand{\degr}{$^{\mathrm{o}}$}
\newcommand{\dPHI}{$\Delta \Phi$\,}

\newenvironment{enumerate*}%
  {\begin{enumerate}%
    \setlength{\itemsep}{0pt}%
    \setlength{\parskip}{0pt}}%
  {\end{enumerate}}

\newcommand{\IFAE}{Institut de F\'isica d'Altes Energies (IFAE),  
	The Barcelona Institute of Science and Technology (BIST), 
	Edifici CN, Universitat Aut\`onoma de Barcelona (UAB), Bellaterra, E-08193 Spain}

\author[]{Machiel Kolstein}
\author[]{Mokhtar Chmeissani}
\author[]{Andreu Pacheco}
\affil[]{\IFAE}

\title{Monitoring Head Movement in a Brain PET Scanner}
\date{}

\begin{document}

\maketitle

\begin{abstract}
When acquiring PET images, body motions are unavoidable, 
	given that the acquisition time could last 10-20 minutes or more. 
These motions can seriously deteriorate the quality of the final image 
	at the level of image reconstruction and attenuation corrections.
Movements can have rhythmic patterns, related to respiratory or cardiac motions, 
	or they can be abrupt reflexive actions caused by the patient's discomfort. 
Many approaches, software and hardware, have been developed to mitigate this problem
	where each approach has its own advantages and disadvantages.
%
%
In this work we present a simulation study of a head monitoring device, named CrowN@22, 
    intended to be used in conjunction with a dedicated brain PET scanner. 
The CrowN@22 device consists of six point sources of non-pure positron emitter isotopes, 
    such as $^{22}$Na or $^{44}$Sc, mounted in crown-like rings around the head of the patient. 
The relative positions of the point sources are predefined and their actual position, 
    once mounted, can be reconstructed by tagging the extra 1274 keV photon (in the case of $^{22}$Na). 
%
These two factors contribute to a superb signal-to-noise ratio, 
	distinguishing between the signal from the $^{22}$Na monitor point sources 
	and the background signal from the FDG in the brain.
%
Hence, even with a low activity for the monitor point sources, 
	as low as 10 kBq per point source, 
	in the presence of 75 MBq activity of $^{18}$F in the brain, 
	one can detect brain movements with a precision of less than 0.3 degrees, or 0.5 mm, 
	which is of the order of the PET spatial resolution, at a sampling rate of 1 Hz.
\end{abstract}
%

%




%
\section{Introduction}

%
Positron Emission Tomography (PET)~\cite{Bailey2005} is a medical imaging modality 
	used for many applications such as oncology, neurology, cardiology, and pharmacology.
Unlike anatomic imaging modalities such as Computed Tomography (CT) and Magnetic Resonance Imaging (MRI), 
    PET provides information of the physiological activity in the body.
%

%
PET can play an important role as an imaging tool for the preclinical diagnosis 
	of neurodegenerative disorders such as Alzheimer's disease (AD)~\cite{Brown2014, Sperling2014}.
For example, the uptake of FDG, a common PET tracer, 
    in parts of the brain where AD affects prematurely such as the posterior cingulate 
    and the precuneus (cortical areas) is studied for early signs of AD~\cite{Choo2007}. 
Moreover, there are promising recent results in the development of imaging biomarkers 
        that bind to the presence of abnormal proteins (amyloid plaques and tau proteins) 
        in the brain~\cite{Brown2014, Sperling2014, Ossenkoppele}.
%

%
%
%
%
%
%

%
Various techniques, at the level of either hardware and software, 
	aim to reach the intrinsic performance of a PET scanner set by physics 
    (\cite{Moses_2011, Levin_1999}) to obtain a high quality PET scan image. 
These efforts are challenged, in clinical PET, by the fact that patients inevitably move during data taking scans 
	which causes significant deterioration to the quality of the final image. 
In PET particularly this is the case due to the long acquisition time of 10-20 minutes or more. 
Whereas some of these movements have a rhythmic pattern due to respiratory or cardiac motions, 
	other movements are abrupt and sudden caused by the patient's reflexive actions due to discomfort. 
Motion correction for various imaging modalities has become 
	an important issue to provide the medical community with an image quality necessary for a proper diagnosis.
Depending on the type of body motion a dedicated algorithm is applied to recover the quality of images. 
In the case of breathing and heartbeats, one can use the gating technique 
    which could be hardware or software driven. 
In the software-driven gating technique, the corrections are derived from the full data set 
	by selecting (gating) the data when the body is in rest (e.g. after exhalation). 
The hardware-driven gating technique is based on the use of a set of external sensors, such as cameras, 
	which collect data, independent of the PET scanner, by monitoring the body movements. 
Later, the data from the sensors provide the information to correct the PET raw data accordingly. 
A review on the techniques used to correct PET images due to body movements can be found in~\cite{Wang_2024}.
In preclinical PET studies (\cite{Miranda2017,Arias2023})
    the movements of a rat's head was tracked using sodium polyacrylate grains loaded with FDG and attached to the head.
%
%

%
%
Although the discussion here could be extended to standard PET scanners, 
	with and without the accompaniment of a CT scan, 
	we restrict our study to the case of dedicated brain PET scanners.
A new generation of dedicated brain PET scanners 
    (\cite{Karp_2003, Akamatsu_2022, Tao_2020, Ahnen_2020, Bauer_2016, Wang_2013, Benlloch_2018, Yamamoto_2011, Jung_2012, Watanabe_2017})
    has emerged to simplify the brain scan procedure by having the patient in a sitting position 
    instead of a supine position. 
The images from such brain PET scanners are more prone to head movements.
In this article, a method is presented to monitor the brain movement during a brain scan. 
The method introduces a device consisting of six point-sources filled with a non-pure positron emitter isotope, 
    specifically $^{22}$Na, arranged in two crown-like bands around the head of the patient. 
The bands can be easily attached to the skull of the head without any additional preparation or selection necessary.
In the following we refer to this device as CrowN@22.
A brain PET scanner in the sitting position is not equipped with a CT scanner.
The CrowN@22 device can be used both in the PET scanner as well as in the independent CT scan for PET-CT co-registration. 
This process is essential for applying the attenuation correction to the final brain PET image.
%
%
The fundamental idea is that events with two 511 keV gammas coming from the $^{22}$Na sources 
    can be easily distinguished from the background (i.e. the two 511 keV coming from FDG in the brain)
    because they come accompanied with a third gamma with energy of 1274 keV with every $^{22}$Na disintegration. 
%
Non-pure isotopes are a convenient method to tag signals and determining the locus of particular regions of the target 
    (\cite{Andreyev2011, Beyene2023, Pratt2023, Chmeissani_2024, GRIGNON2007142, Conti_2016, Kolstein_2016, SITARZ2020108898}).   
By using non-pure PET tracers such as $^{22}$Na, the SNR is far better than when using pure PET tracers, as used in the brain, 
    taking advantage of selecting the signal by means of tagging triple coincidences. 
This increases the SNR significantly by comparison to work reported in \cite{Miranda2017,Arias2023}, 
    even in a challenging environment where the scattering fraction could be as high as 40\%.
Movements of the head will change the detected positions of the $^{22}$Na sources 
	and the PET signals of the brain can be realigned accordingly.
The performance of the device has been studied by a full simulation of the setup, 
	as explained in the following sections.

\section{Materials and Methods}

\subsection{Simulation of a brain PET scintillation scanner}

\begin{figure}[th!]
    \centering
    \begin{tabular}{l c r}
    \includegraphics[width=0.25\linewidth]{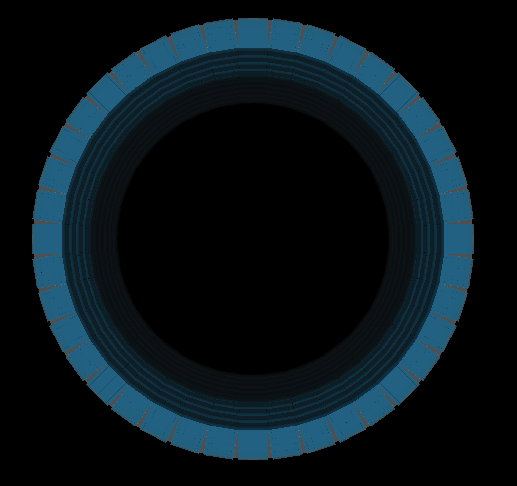}
      &
    \includegraphics[width=0.25\linewidth]{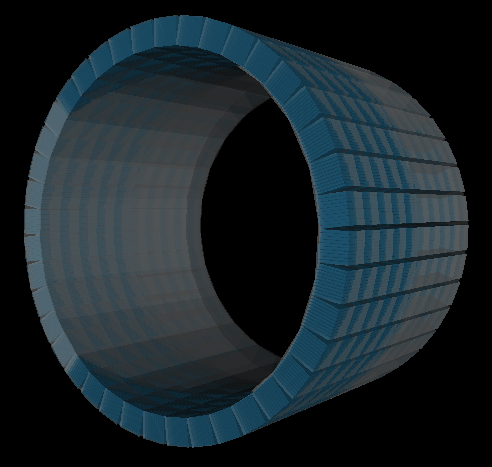}
      &      
    \includegraphics[width=0.27\linewidth]{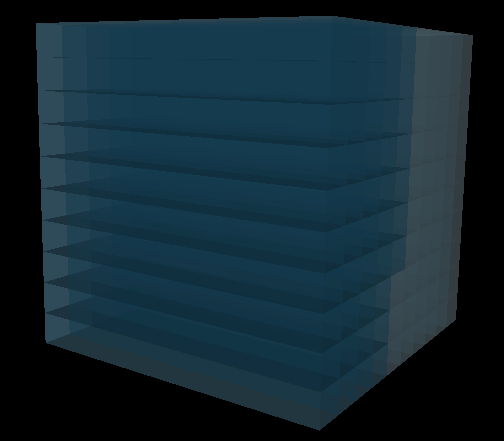} 
    \end{tabular}
\caption{3D depiction of the PET scintillator scanner geometry used in the simulations for this article.
Left and centre: the entire scanner.
Right: one block of $10 \times 10$ scintillation LYSO crystals.}
\label{fig:geometry}
\end{figure}

For the simulation of the scanner, the phantoms and the physics processes, the Geant4 based software package 
GAMOS (Geant4-based Architecture for Medicine-Oriented Simulations) is used~\cite{GAMOS}.
\begin{table}[th!]
\caption{Summary of properties of the simulated PET scanner}
\label{tab:simu_scanner_properties}
\begin{tabular}{|l|l|}
\hline
Detector material    & LYSO crystals        \\ \hline
Crystal size         & 2 $\times$ 2 $\times$ 20 mm$^3$       \\ \hline
Crystal energy resolution         & 10\% FWHM at 511 keV \\ \hline
Scanner inner radius & 127 mm               \\ \hline
Scanner axial size   & 200 mm               \\ \hline
\end{tabular}
\end{table}
The brain PET scintillator scanner used in the simulation is depicted in figure~\ref{fig:geometry} 
and its properties are summarised in table~\ref{tab:simu_scanner_properties}.
The scanner consists of 10 scanner rings in the axial direction, where 
    each ring has an inner radius of 127 mm  
    and is built up out of 40 blocks in the circular direction.
Each block is an array of $10 \times 10$ scintillation LYSO crystals
    where each crystal has a radial length of 20 mm and a transversal area of $2 \times 2$ mm$^2$.
Hence, the total axial size of the scanner is 200 mm, from z = -100 mm to z = + 100 mm.
The energy resolution of the LYSO crystals is simulated as 10\% FWHM at 511 keV.

%
%
The basic parameters of the simulated PET scanner are summarized in table ~\ref{tab:simu_scanner_properties} 
    and are meant to be generic.  
The aforementioned parameters are comparable to those of dedicated brain PET scanners 
    (\cite{Karp_2003, Ahnen_2020, Yamamoto_2011, Jung_2012, Watanabe_2017, Won_2021, Yoshida_4179539}), 
    where the energy resolutions varies from 12\% to 24\% 
    and the crystal size in varies between 1.2 $\times$ 1.2 $\times$ 20 mm$^3$
    to 4.9 $\times$ 5.9 $\times$ 7.5 mm$^3$. 
%
%

\subsection{Monitoring head movement}

\begin{figure}[th!]
    \centering
    \includegraphics[width=0.70\linewidth]{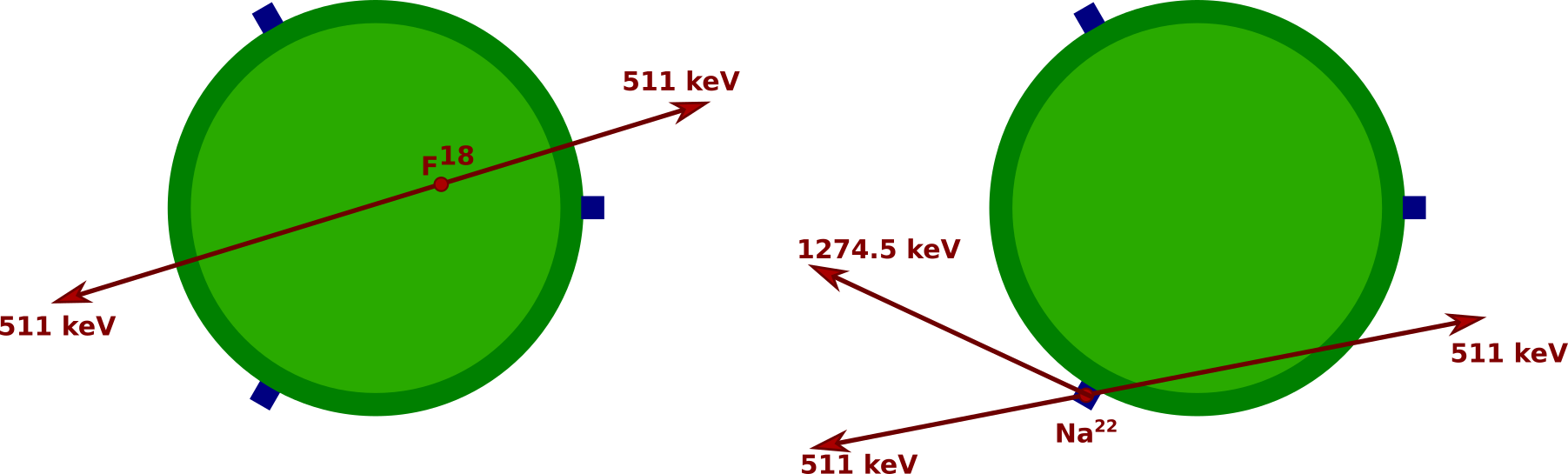}
    \caption{Depiction of standard coincidences and triple coincidences in one of the two monitor rings around the head.
Left: a standard coincidence of two 511 keV gammas, coming from the positron emission of the $^{18}$F radiotracer in the brain.
Right: a triple coincidence of two 511 keV gammas plus an additional 1274 keV of the $^{22}$Na radiotracer in the monitor point sources.}
\label{fig:depictingTechnique} 
\end{figure}

\begin{figure}[th!]
    \centering
    \begin{tabular}{l c r}
    \includegraphics[width=0.20\linewidth]{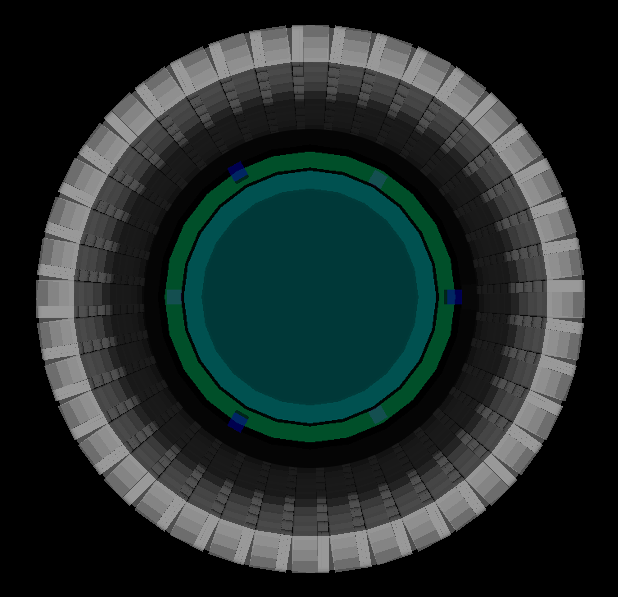}    
    &
    \includegraphics[width=0.42\linewidth]{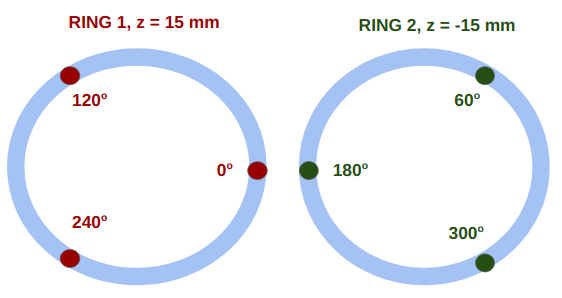}
    \end{tabular}
\caption{Depiction of the $^{22}$Na monitor point sources in two rings.
Left: Front view of the $^{22}$Na point sources arranged in two rings 
       and the PET scanner (grey), the simulated brain (light blue) and cranium (turquoise). 
       The front ring (green) with 3 $^{22}$Na point sources (dark blue) can be clearly seen. 
Centre and right: Depiction of the location of the 3 front and the 3 back point sources.
}
\label{fig:monitors2rings}
\end{figure}

The presence of a head inside the PET scanner is simulated with a cylinder of radius 80 mm and length 70 mm 
    filled with simulated brain material and an FDG ($^{18}$F) radiotracer.
The cylinder is surrounded by 10 mm thick bone material to simulate the cranium.
To monitor the movement of the head during a PET scan, 
    6 point sources were used, dosed with the $^{22}$Na isotope
    and placed in 2 crown-like rings, with 3 point sources each, around the head.
The $^{22}$Na isotope emits both a positron and an additional gamma particle with an energy of 1274 keV.
Whereas usually in standard PET, using a FDG radiotracer, one collects coincidences between two 511 keV gammas, 
    now, with a $^{22}$Na source, triple coincidences of two 511 keV gammas plus the additional 1274 keV gamma
    can be collected.
The difference between the two types of coincidences is shown in figure~\ref{fig:depictingTechnique}.
As usual, the energy detections of the two 511 keV gammas in the scanner are used to construct Lines Of Response (LOR), 
    which serves as input for the image reconstruction. 
The detection of the additional 1274 keV gamma is used to tag the triple coincidences and distinguish them from the 
    standard two gamma coincidences coming from the FDG radiotracer located in the brain phantom.
The triple coincidences are used to exclusively create an image of the monitoring point sources. 
By detecting movement of the monitor sources in subsequent image sets, 
    one can monitor the movement of the brain and correct for it.
The $^{22}$Na monitor point sources were arranged in two different rings with radius 95 mm, with 3 monitors per ring 
and a distance of 30 mm between each other (figure~\ref{fig:monitors2rings}).


\subsubsection{Event selection, image reconstruction and image analysis}

The output of the GAMOS simulation is converted into a time-ordered list of energy depositions in the scanner crystals. 
Subsequently, triple coincidences are selected of two 511 keV gammas and the additional 1274 keV gamma. 
Triple coincidences are accepted when exactly three hits are detected within a coincidence time window of 1 ns, 
 where two hits are required to have energy depositions of 511 keV $\pm$ 40 keV 
 and the third hit is required to have an energy deposition of 1274.0 $\pm$ 75 keV.
The LORs from the two 511 keV gammas are the input for the image reconstruction, 
    using an implementation of the iterative LM-OSEM algorithm developed by the authors~\cite{Kolstein_2014}.
An evaluation of this implementation and comparison with other image reconstruction algorithms 
    can be found in~\cite{Mikhaylova2014}.
The resulting images are analysed with the AMIDE software~\cite{AMIDE_article, AMIDE_web}, 
    extracting the weighted centres of the regions of interest around the $^{22}$Na monitor point sources.
%
%
Hospitals have access to professional and sophisticated software to establish regions of interest in an image.
However, for the purpose of proving the validity of the CrowN@22 concept it is not necessary to develop such software.

\subsubsection{Fitting three-dimensional planes through a set of points}

Once we have constructed an image from the LORs that were tagged with the additional 1274 keV gamma 
    emitted by the $^{22}$Na monitor point sources, we can identify the point sources in the image.
For the study as presented in this article, we used regions of interest (ROIs) to establish the weighted centre position of each monitor point source.
Subsequently, a program is used to fit planes through the 3 weighted centre positions of each of the 2 rings.
%
%
%
The fitting program is based on a Singular Value Decomposition (SVD) algorithm~\cite{fitting_program} 
    and calculates the circle centre, radius and the normal vector perpendicular to the circle plane. 
The inner product between the normal to the plane in the case of a rotated head 
    and the normal to a reference plane where the head was not rotated
    provides the polar angle $\Theta$. 
%
%

\subsubsection{Rotation angles used in the simulations}

\begin{figure}[th!]
\centering
    \includegraphics[width=0.20\linewidth]{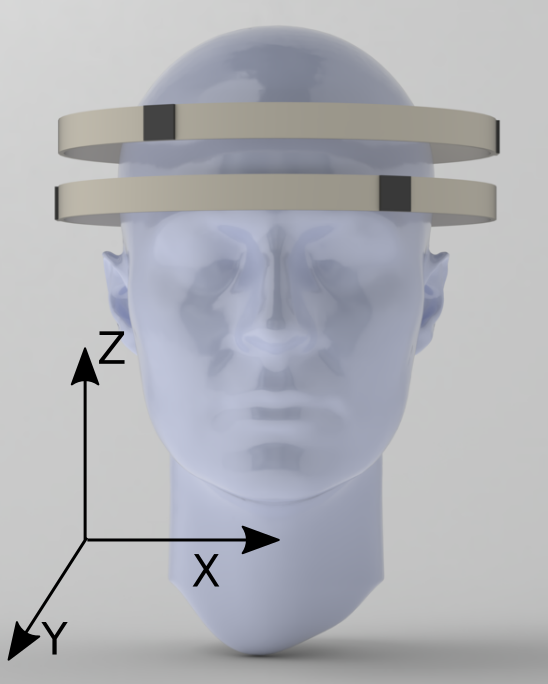}
\caption{Illustration of the position of the head inside the coordinate system.}
\label{fig:head_in_space} 
\end{figure}

All rotations in our simulations are done in a particular order, with respect to the global coordinate system.
The location of the head with respect to the global coordinate system is illustrated in figure~\ref{fig:head_in_space}.
First, a rotation around the z-axis corresponds to a movement of the head looking to the left or right.
Without any rotation in the x-y frame, the point sources are already distributed around the circle with different angles $\Phi$, as shown in figure~\ref{fig:monitors2rings}. 
Any rotation around the z-axis of this set-up is denoted in this article with \dPHI.
Subsequently, the monitor rings are rotated around the y-axis, which we will denote with the angle $\Theta(y)$.
This corresponds to a movement where the ear is moved to the shoulder (left or right).
Finally, the resulting rings can be rotated around the x-axis, which we will denote with the angle $\Theta(x)$. 
This corresponds to a movement where the head nods yes (up or down).
The analysis of the simulation data starts with the reverse situation where 
    we have the weighted centre positions of all 6 point sources.
The calculation of the angles from our data is done in a few steps for each ring: 
\begin{enumerate*}
    \item Fit a plane through the source positions and get the normal vector with respect to the plane. 
    \item Project the normal vector onto the y-z plane and get $\Theta(x)$ from the dot product of the projected normal and the z-axis.
    \item Rotate the normal vector back over $-\Theta(x)$.
    \item Project the back-rotated normal vector onto the x-z plane and get $\Theta(y)$ from the dot product of the projected normal and the z-axis.
    \item Rotate the source positions back first over  $-\Theta(x)$ (around the x axis) and then over $-\Theta(y)$ (around the y axis).
    \item Get $\Delta \Phi$ from $\arctan(y/x)$ for each of the back-rotated source positions.
\end{enumerate*}
This way, we can compare our results for $\Delta \Phi$, $\Theta(x)$ and $\Theta(y)$ exactly with the corresponding values used in the original GAMOS data generation file.
This procedure can be easily generalised into a procedure using standard polar coordinates without loss of generality. 


\subsection{Overview of measurements}

%
%
In section~\ref{sec:with_phantom}, 
    a brain phantom filled with the $^{18}$F isotope, with an activity of 75 MBq, 
    is simulated in addition to the monitor ring, where each $^{22}$Na point source has an activity of 100 kBq.
The purpose of this section is to show that with relative low activities 
    for the $^{22}$Na point sources, 
    the high background signal from the brain phantom does not affect the monitor signal, 
    thanks to the triple coincidence tagging.
The section shows relative coincidence count rates and images for both type of coincidences.
In the remaining sections no brain phantom was used in order to speed up the simulation. 
In all cases, enough events are simulated to provide for 10 independent sets of 500 triple coincidences each. 
Unless explicitly stated otherwise, in all cases the crystal size was $2 \times 2 \times 20$ mm$^3$ 
    and the energy resolution was 10\% FWHM.

In section~\ref{sec:without_phantom_scanThetaY} various rotations around the y-axis are performed 
    and for each angle the resulting measurements are shown. 

In section~\ref{sec:without_phantom_rotXYZ} a particular combination 
    of rotations around the x-, y- and z-axis is tested, with \dPHI = 30\degr, $\Theta(x) = 20$\degr \ and 
    $\Theta(y) = 10$\degr.
In section~\ref{sec:without_phantom_rotXYZ_dEE}  the same combination of rotations is tested
    for a different crystal energy resolution of 15 \% FWHM.
In section~\ref{sec:without_phantom_rotXYZ_size} the same combination of rotations is tested 
    for a different crystal size of $3 \times 3 \times 20$ mm$^3$

In section~\ref{sec:without_phantom_rotXYZ_shiftZ} the same combination of rotations is tested 
    with additional shifts in the axial direction of 3 mm, 6 mm or 10 mm.

In section~\ref{sec:without_different_PHIs}, the position of the point sources along the rings is changed, 
    resulting in a less symmetric arrangement.
Additionally, the rings are rotated with respect to the global coordinate system with $\Theta(x) = 20$\degr \ and 
    $\Theta(y) = 10$\degr.

\section{Results}

\label{sec:Results}


%
%

\subsection{With brain phantom.}

\label{sec:with_phantom}

\begin{figure}[th!]
    \centering
    \begin{tabular}{l r}
    \includegraphics[width=0.31\linewidth]{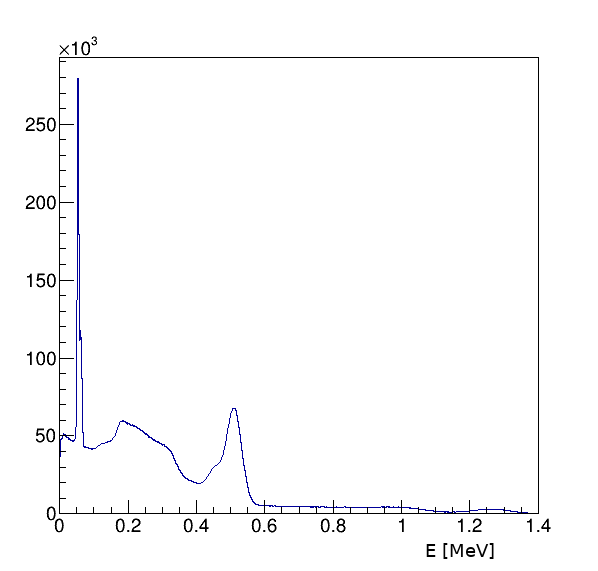}
      &
    \includegraphics[width=0.31\linewidth]{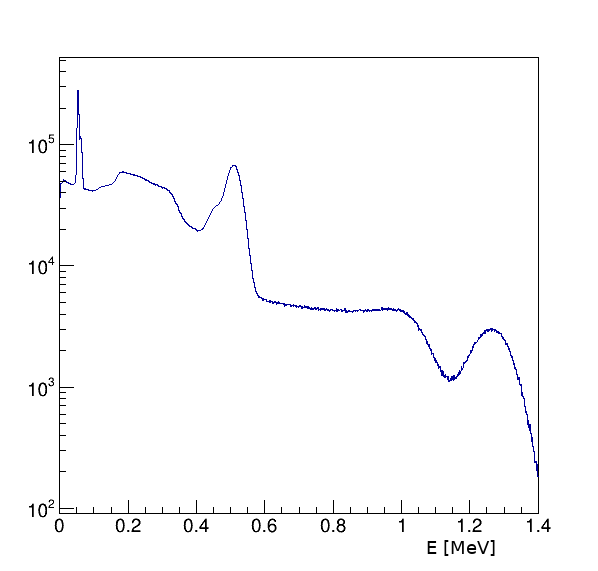}
    \end{tabular}
\caption{
Left: energy distribution of the single hits in the scanner. 
Right: logarithmic scale of the same energy distribution to show the additional 1274 keV gamma contribution.}
\label{fig:NONROT_LM}
\end{figure}
\begin{figure}[th!]
    \centering
    \begin{tabular}{l c r}
    \includegraphics[width=0.2\linewidth]{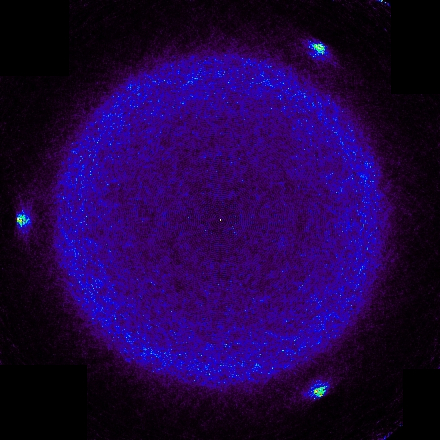}
      & 
    \includegraphics[width=0.2\linewidth]{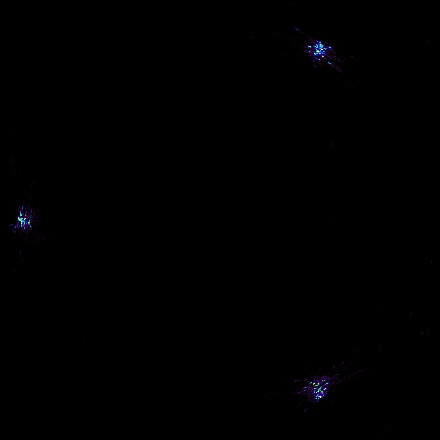}
      & 
    \includegraphics[width=0.2\linewidth]{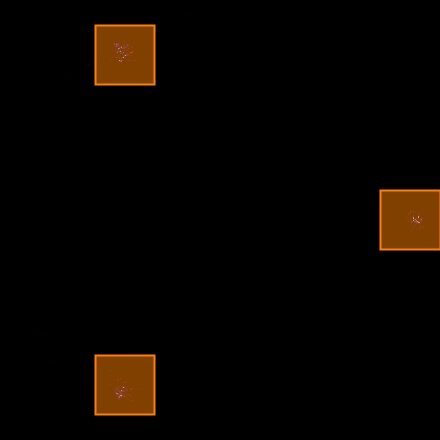}
    \end{tabular}
\caption{Left: final image for the set of standard double coincidences at the slice corresponding to the front ring position.
Centre: final image for the set of triple coincidences at the same slice.
Right: The Regions of Interest (ROIs) superimposed onto the image from triple coincidences.}
\label{fig:bg_NONROT_AMIDE_frontring}
\end{figure}
\begin{figure}
    \centering
    \begin{tabular}{c}
    \includegraphics[width=0.69\linewidth]{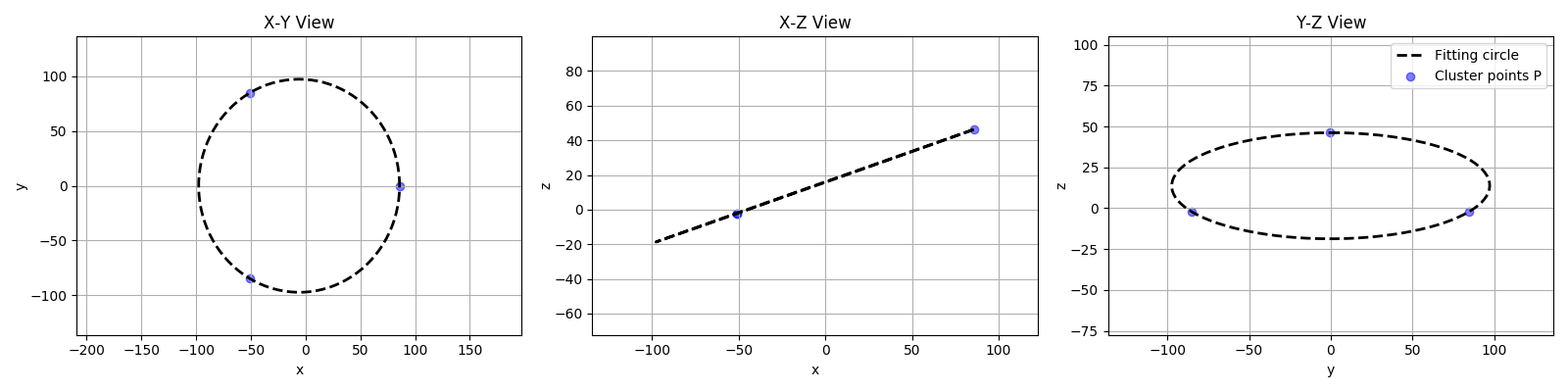}
    \end{tabular}
    \caption{Visualisation of a fit to the positions of the monitor point sources on a ring, 
        with $\Theta = 20^{\circ}$, projected in the xy, xz and yz frames.}
    \label{fig:10x5k_meshFigs}
\end{figure}

The first group of simulations were done with the inclusion of the brain phantom filled with $^{18}$F
    as illustrated in figure~\ref{fig:monitors2rings} for three different $\Theta(y)$ angles. 
Figure~\ref{fig:NONROT_LM} shows the energy distribution of the single hits in the PET scanner. 
The total activity for all 6 $^{22}$Na monitors was set to 600 kBq 
    while the total activity for the phantom cylinder was set to 75 MBq.
Data was taken with a non-rotated brain, with $\Theta(y)$ equal to 10$^\mathrm{o}$ 
    and $\Theta(y)$ equal to 20$^\mathrm{o}$.
Figure~\ref{fig:bg_NONROT_AMIDE_frontring} shows what the final images look like using either double coincidences 
    or triple coincidences.
Table~\ref{tab:phantom_coinc_efficiencies} shows the number of either standard PET coincidences 
    or triple coincidences found. 
\begin{table}
\caption{After generating 20 $\cdot 10^6$ isotope decays (either from $^{18}$F in the brain phantom or from $^{22}$Na in the monitor point sources), this table shows the number of standard and triple coincidences. For the analysis of the head movement from the position of the monitor point sources, triple coincidences are used. }
\begin{tabular}{|r|r|r|}
\hline
\textbf{True $\Theta(y)$}  & \begin{tabular}[c]{@{}l@{}} \textbf{Standard PET coincidences}
    \\ ($ 2 \times 511 \pm 40$ keV)\end{tabular} 
            & \begin{tabular}[c]{@{}l@{}} \textbf{Triple coincidences}
    \\ ($2 \times 511 \pm 40$ keV $+ 1274.0 \pm 75.0$ keV)\end{tabular} \\ \hline
\textbf{0\degr}  & 956482    & 1535   \\ \hline
\textbf{10\degr} & 966418    & 1555   \\ \hline
\textbf{20\degr} & 988875    & 1466   \\ \hline
\end{tabular}
\label{tab:phantom_coinc_efficiencies}
\end{table}
For each $\Theta(y)$, the sample of triple coincidences for all 6 monitors 
    was divided into sets of 150, 300, or 500 total coincidences. 
%
Table~\ref{tab:phantom_multitheta_multicoinc_results} shows that 
    even for a relatively small number of total triple coincidences, 
    we get consistent results for the estimation of $\Theta(y)$.
Figure~\ref{fig:10x5k_meshFigs} shows an example of a fit through a plane of 3 point sources in case $\Theta = 20^\mathrm{o}$.
\begin{table}
\caption{Mean estimated $\Theta(y)$ and root mean square (RMS) for different $\Theta(y)$ angles 
    and different number of total monitor coincidences, for data with an additional $^{18}$F background phantom.
The means were calculated for 10 sets of 150 coincidences, 5 sets of 300 coincidences, 
    and 3 sets of 500 coincidences.
}
\begin{tabular}{|r|r|r|r|}
\hline
\textbf{True $\Theta(y)$}  
                 & \textbf{10 sets} $\mathbf{\#C = 150}$ 
                 & \textbf{5 sets} $\mathbf{\#C = 300}$ 
                 & \textbf{3 sets} $\mathbf{\#C = 500}$ \\ \hline 
0\degr  &   0.2$ \pm $0.2\degr         &   0.1$ \pm $0.1\degr          &  0.1$ \pm $0.1\degr   \\ 
10\degr &   9.5$ \pm $0.2\degr         &   9.5$ \pm $0.1\degr          &  9.5$ \pm $0.1\degr   \\ 
20\degr &  19.4$ \pm $0.3\degr         &  19.4$ \pm $0.2\degr          & 19.4$ \pm $0.1\degr  \\ \hline
\end{tabular}
\label{tab:phantom_multitheta_multicoinc_results}
\end{table}
The $\Phi$ angles for monitor point sources are 0\degr, 120\degr \ and 240\degr \ on the front monitor ring
    and 60\degr, 180\degr \ and 300\degr \ on the back monitor ring, 
    where $\Phi$ is defined as the angle in the x-y coordinate frame.
If we define \dPHI as any subsequent rotation in the x-y frame, we expect \dPHI to be 0 in the current case.
Table~\ref{tab:phantom_multiphi} gives an overview of the measured values for \dPHI, for different values of $\Theta(y)$.

\begin{table}[th!]
\caption{Mean estimated deviation of \dPHI and RMS for different $\Theta(y)$ angles, 
    averaged over 10 sets of 500 coincidences. 
    The rows denoted by <arc length> show the average over 6 monitor point sources.  } 
\begin{tabular}{|l|l|l|l|l|l|l|}
\hline
\textbf{ $\Theta$(y) = 0\degr }
            &   \multicolumn{6}{l|}{\textbf{True \dPHI} = 0.0\degr }     \\ \hline 
\textbf{measured}   
             & -0.2$ \pm $0.4\degr & -0.1$ \pm $0.2\degr & 0.1$ \pm $0.1\degr  
             & -0.2$ \pm $0.2\degr & -0.1$ \pm $0.2\degr & 0.1$ \pm $0.2\degr   \\ \cline{1-7} 
\textbf{arc length [mm]}  
             & -0.3$ \pm $0.7 & -0.2$ \pm $0.3 & 0.2$ \pm $0.1  
             & -0.3$ \pm $0.3 & -0.2$ \pm $0.2 & 0.2$ \pm $0.4  \\ \hline
\textbf{<arc length> }            
        & \multicolumn{6}{l|}{-0.1$ \pm $0.4 mm}  \\ \hline 
             
\textbf{ $\Theta$(y) = 10\degr }
            &   \multicolumn{6}{l|}{\textbf{True \dPHI} = 0.0\degr }     \\ \hline
\textbf{measured}       
            & 0.2$ \pm $0.2\degr  & 0.1$ \pm $0.2\degr  & -0.2$ \pm $0.3\degr 
            & -0.1$ \pm $0.3\degr & -0.1$ \pm $0.1\degr & -0.1$ \pm $0.2\degr \\ \cline{1-7} 
\textbf{arc length [mm]}    
            & 0.3$ \pm $0.3  & 0.2$ \pm $0.2  & -0.3$ \pm $0.5 
            & -0.1$ \pm $0.5 & -0.2$ \pm $0.2 & -0.2$ \pm $0.3   \\ \hline
\textbf{<arc length>  }            
        & \multicolumn{6}{l|}{0.1$ \pm $0.4 mm}  \\ \hline 
            
\textbf{ $\Theta$(y) = 20\degr }
            &   \multicolumn{6}{l|}{\textbf{True \dPHI} = 0.0\degr }     \\ \hline            
\textbf{measured}      
            & -0.2$ \pm $0.2\degr & 0.1$ \pm $0.1\degr  & 0.0$ \pm $0.1\degr   
            & 0.1$ \pm $0.3\degr  & 0.0$ \pm $0.2\degr  & -0.1$ \pm $0.2\degr \\ \cline{1-7} 
\textbf{arc length [mm]}   
            & -0.4$ \pm $0.3 & 0.2$ \pm $0.2  & 0.0$ \pm $0.2   
            & 0.2$ \pm $0.4  & 0.0$ \pm $0.3  & -0.2$ \pm $0.3 \\ \hline
\textbf{<arc length>  }            
        & \multicolumn{6}{l|}{0.0$ \pm $0.3 mm}  \\ \hline 
        
\end{tabular}
\label{tab:phantom_multiphi}
\end{table}

%
%

\subsection{Scanning rotation around y-axis}

\label{sec:without_phantom_scanThetaY}

\begin{figure}[th!]
    \centering
    \begin{tabular}{l r}    
    \includegraphics[width=0.62\linewidth]{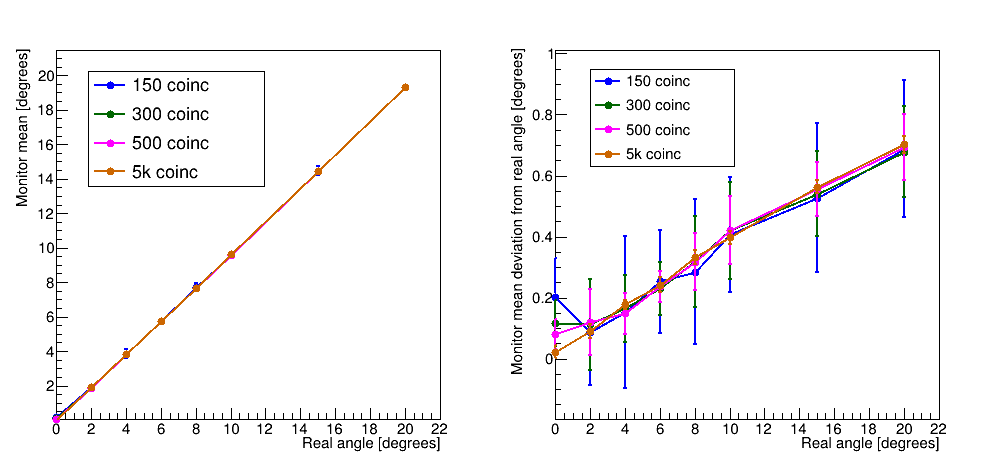}
         &
    \includegraphics[width=0.29\linewidth]{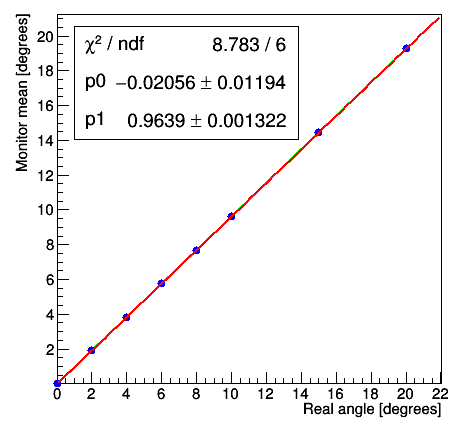}
    \end{tabular}
    \caption{
    Left: measured $\Theta(y)$ angle versus the true $\Theta(y)$ for different number of total monitor coincidences.
    Centre: deviation from the true $\Theta(y)$ angle versus the true $\Theta(y)$ for different number of total monitor coincidences.
    Right: Fit to the measured $\Theta(y)$ angle versus the true $\Theta(y)$ for the case of 5k coincidences.
    }
    \label{fig:multitheta_multicoinc_results}
\end{figure}

Next, simulations were done without using an additional brain phantom, 
    in order to speed up the simulation and generate enough events to get multiple sets of triple coincidences.
Data was generated for eight different $\Theta(y)$ angles 
    and in each case 10 sets of 150, 300 and 500 total triple coincidences were collected
The results are shown in table~\ref{tab:multitheta_multicoinc_results}.
%
%
The relation between the measured angle and the true angle, for different number of triple coincidences, 
    is shown in figure~\ref{fig:multitheta_multicoinc_results}.
\begin{table}
\caption{Mean estimated $\Theta(y)$ and RMS for different true $\Theta(y)$ angles 
    and different number of total triple coincidences. 
    No background phantom was simulated.}
\begin{tabular}{|r|r|r|r|r|}
\hline
\textbf{True $\Theta(y)$}  
            & \textbf{\#C = 150}  & \textbf{\#C = 300} & \textbf{\#C = 500} & \textbf{\#C = 5000 } \\ \hline
0\degr  & 0.2$ \pm $0.1\degr      & 0.1$ \pm $0.1\degr      & 0.1$ \pm $0.1\degr      & 0.02$ \pm $0.02\degr  \\ 
2\degr  & 1.9$ \pm $0.2\degr      & 1.9$ \pm $0.2\degr      & 1.9$ \pm $0.1\degr      & 1.91$ \pm $0.02\degr  \\
4\degr  & 3.9$ \pm $0.3\degr      & 3.8$ \pm $0.1\degr      & 3.9$ \pm $0.1\degr      & 3.82$ \pm $0.02\degr  \\
6\degr  & 5.8$ \pm $0.2\degr      & 5.8$ \pm $0.1\degr      & 5.8$ \pm $0.1\degr      & 5.76$ \pm $0.02\degr  \\
8\degr  & 7.7$ \pm $0.2\degr      & 7.7$ \pm $0.2\degr      & 7.7$ \pm $0.1\degr      & 7.67$ \pm $0.02\degr  \\
10\degr & 9.6$ \pm $0.2\degr      & 9.6$ \pm $0.2\degr      & 9.6$ \pm $0.1\degr      & 9.60$ \pm $0.02\degr  \\
15\degr & 14.5$ \pm $0.2\degr     & 14.4$ \pm $0.1\degr     & 14.4$ \pm $0.1\degr     & 14.44$ \pm $0.02\degr \\
20\degr & 19.3$ \pm $0.2\degr     & 19.3$ \pm $0.1\degr     & 19.3$ \pm $0.1\degr     & 19.30$ \pm $0.03\degr  \\ \hline
\end{tabular}
\label{tab:multitheta_multicoinc_results}
\end{table}
%
%
%
%

%
Since the deviation from the true $\Theta(y)$ shows a linear behaviour as a function of $\Theta(y)$, 
    it can be corrected for.
Using the results for 5k coincidences, 
        as shown in table~\ref{tab:multitheta_multicoinc_results}, 
	we can do a polynomial fit of degree 1 of the form $f(x) = p_0 + p_1 \cdot x$.
The right plot in figure~\ref{fig:multitheta_multicoinc_results} 
	shows that from the fit we obtain $p_0 = -0.021$ and $p_1 = 0.964$.
Applying the correction $\Theta^\ast = (\Theta - p_0)/p_1$  
    to the values in table~\ref{tab:multitheta_multicoinc_results} 
    for the cases of 150, 300 and 500 coincidences, 
    gives the values shown in table~\ref{tab:multitheta_multicoinc_corrected}.
The corrected values $\Theta^\ast(y)$ can be translated into arc lengths 
    with $A = \Theta^\ast \cdot \frac{\pi}{180} \cdot R$, with R = 95 mm.
Applying this to the deviations from the true values
    given in table~\ref{tab:multitheta_multicoinc_results}, 
    will give the deviation in arc lengths, shown in table~\ref{tab:corrected_arc_lengths}.
This shows that even with only 150 total triple coincidences
    we can get very accurate values for $\Theta(y)$, with errors of the arc length of less than 0.5 mm.
\begin{table}[th!]
\caption{Mean estimated $\Theta^\ast$ and difference with true values
    after applying a correction curve from the fit of figure~\ref{fig:multitheta_multicoinc_results}
    for different number of total monitor coincidences.}
\begin{tabular}{|r|r|r|r|r|r|r|}
\hline
    & \multicolumn{2}{l|}{\textbf{\#C = 150}}
    & \multicolumn{2}{l|}{\textbf{\#C = 300}} 
    & \multicolumn{2}{l|}{\textbf{\#C = 500}} \\ \hline
\textbf{True $\Theta(y)$}     
     & \multicolumn{1}{l|}{\textbf{Experiment}} & \textbf{True - exp.} 
     & \multicolumn{1}{l|}{\textbf{Exp.}} & \textbf{True - exp.}
     & \multicolumn{1}{l|}{\textbf{Exp.}} & \textbf{True - exp.} \\ \hline
0\degr         & 0.2$ \pm $0.1\degr  &  -0.2$ \pm $0.1\degr   & 0.1$ \pm $0.1\degr  & -0.1$ \pm $0.1\degr
          & 0.1$ \pm $0.1\degr  &  0.1$ \pm $0.1\degr  \\ \hline
2\degr         & 2.0$ \pm $0.2\degr  &   0.0$ \pm $0.2\degr   & 2.0$ \pm $0.2\degr  &  0.0$ \pm $0.2\degr
          & 2.0$ \pm $0.1\degr  &  0.0$ \pm $0.1\degr  \\ \hline
4\degr         & 4.0$ \pm $0.3\degr  &  0.0$ \pm $0.3\degr   & 4.0$ \pm $0.1\degr   &  0.0$ \pm $0.1\degr
          & 4.0$ \pm $0.1\degr  &  0.0$ \pm $0.1\degr   \\ \hline
6\degr         & 6.0$ \pm $0.2\degr  &  0.0$ \pm $0.2\degr   & 6.0$ \pm $0.1\degr   & 0.0$ \pm $0.1\degr
          & 6.0$ \pm $0.1\degr  &  0.0$ \pm $0.1\degr  \\ \hline
8\degr         & 8.0$ \pm $0.2\degr  &  0.0$ \pm $0.2\degr   & 8.0$ \pm $0.2\degr   &  0.0$ \pm $0.2\degr   
          & 8.0$ \pm $0.1\degr  &  0.0$ \pm $0.1\degr  \\ \hline
10\degr        & 10.0$ \pm $0.2\degr  &  0.0$ \pm $0.2\degr   & 10.0$ \pm $0.2\degr  &  0.0$ \pm $0.2\degr   
          & 10.0$ \pm $0.1\degr  &  0.0$ \pm $0.1\degr  \\ \hline
15\degr        & 15.0$ \pm $0.2\degr &  0.0$ \pm $0.2\degr   & 15.0$ \pm $0.1\degr & 0.0$ \pm $0.1\degr   
          & 15.0$ \pm $0.1\degr & 0.0$ \pm $0.1\degr    \\ \hline
20\degr        & 20.0$ \pm $0.2\degr &  -0.1$ \pm $0.2        & 20.1$ \pm $0.2\degr & -0.1$ \pm $0.2\degr   
          & 20.1$ \pm $0.1\degr & -0.1$ \pm $0.1\degr    \\ \hline
\end{tabular}
\label{tab:multitheta_multicoinc_corrected}
\end{table}
\begin{table}
\caption{  
Difference between true and experimental values $\Theta^\ast(y)$ after correction
    (from table~\ref{tab:multitheta_multicoinc_corrected}), 
    translated into arc lengths for different number of coincidences.}
\begin{tabular}{|r|r|r|r|}
\hline
        & \textbf{\#C = 150}        
        & \textbf{\#C = 300}       
        & \textbf{\#C = 500}        \\ \hline
\textbf{arc length {[}mm{]}} 
        & \textbf{True - exp.}      
        & \textbf{True - exp.}      
        & \textbf{True - exp.}      \\ \hline
0.0    & {-0.4$ \pm $0.2}  & {-0.2$ \pm $0.1}  & {0.2$ \pm $0.1}  \\ \hline
3.31    & { 0.0$ \pm $0.3}  & {0.0$ \pm $0.3}   & {0.1$ \pm $0.2}  \\ \hline
6.63    & { 0.0$ \pm $0.4}  & {0.0$ \pm $0.2}   & {0.0$ \pm $0.1}  \\ \hline
9.94    & { 0.0$ \pm $0.0}  & {0.0$ \pm $0.2}  & {0.0$ \pm $0.1}  \\ \hline
13.26   & {-0.1$ \pm $0.4}  & {0.0$ \pm $0.3}   & {0.0$ \pm $0.2} \\ \hline
16.57   & {-0.1$ \pm $0.3}  & {0.1$ \pm $0.3}   & {0.1$ \pm $0.2}  \\ \hline
24.86   & {-0.1$ \pm $0.4}  & {0.0$ \pm $0.2}  & {0.0$ \pm $0.2}  \\ \hline
33.14   & {-0.1$ \pm $0.4}  & {-0.1$ \pm $0.3}  & {-0.1$ \pm $0.2}  \\ \hline
\end{tabular}
\label{tab:corrected_arc_lengths}
\end{table}
If we define \dPHI as any rotation of the point sources in the x-y frame 
    with respect to their standard positions, we expect \dPHI to be 0.0 in the current case.
Table~\ref{tab:nophantom_thetayscan_multiphi} gives an overview of the measured values for \dPHI, 
    for different values for the $\Theta(y)$ angle.
\begin{table}[th!]
\caption{Mean estimated deviation of \dPHI and RMS for different $\Theta$ angles, 
    averaged over 10 sets of 500 coincidences.
    The rows denoted by <arc length> show the average over 6 monitor point sources.}
\begin{tabular}{|l|l|l|l|l|l|l|}
\hline
\textbf{ $\Theta$(y) = 0}
   & \multicolumn{6}{l|}{\textbf{True \dPHI} = 0.0\degr }     \\ \hline
\textbf{measured}        
    & 0.0$ \pm $0.1\degr    & -0.1$ \pm $0.2\degr   & -0.1$ \pm $0.3\degr   
    & 0.0$ \pm $0.2\degr    & 0.0$ \pm $0.1\degr    & -0.1$ \pm $0.2\degr   \\ \cline{1-7} 
\textbf{arc length [mm]} 
    & 0.1$ \pm $0.2  & -0.1$ \pm $0.3 & -0.1$ \pm $0.5 
    & 0.1$ \pm $0.3  & 0.03$ \pm $0.2 & -0.1$ \pm $0.3 \\ \hline
\textbf{<arc length>}            
        & \multicolumn{6}{l|}{0.0$\pm$0.3 mm}  \\ \hline 
    
\textbf{ $\Theta$(y) = 10}
   & \multicolumn{6}{l|}{\textbf{True \dPHI} = 0.0\degr }     \\ \hline
\textbf{measured}        
    & 0.0$ \pm $0.3\degr  & -0.1$ \pm $0.1\degr & 0.0$ \pm $0.1\degr  
    & 0.0$ \pm $0.2\degr  & 0.1$ \pm $0.2\degr  & 0.1$ \pm $0.2\degr  \\ \cline{1-7} 
\textbf{arc length [mm]} 
    & 0.0$ \pm $0.4  & -0.2$ \pm $0.2 & 0.0$ \pm $0.2
    & 0.0$ \pm $0.3 &   0.1$ \pm $0.3 & 0.1$ \pm $0.3  \\ \hline
\textbf{<arc length>}            
        & \multicolumn{6}{l|}{0.0$ \pm $0.3 mm}  \\ \hline 
    
\textbf{ $\Theta$(y) = 20}
   & \multicolumn{6}{l|}{\textbf{True \dPHI} = 0.0\degr }     \\ \hline
\textbf{measured}        
    & -0.1$ \pm $0.3\degr  & 0.1$ \pm $0.1\degr  & -0.1$ \pm $0.2\degr 
    &  0.0$ \pm $0.2\degr  & 0.0$ \pm $0.2\degr  &  0.0$ \pm $0.1\degr   \\ \cline{1-7} 
\textbf{arc length [mm]} 
    & -0.2$ \pm $0.6   & 0.2$ \pm $0.2  & -0.2$ \pm $0.3 
    &  0.0$ \pm $0.3  &  0.0$ \pm $0.3  &  0.0$ \pm $0.2   \\ \hline
\textbf{<arc length>}            
        & \multicolumn{6}{l|}{0.0$ \pm $0.3 mm}  \\ \hline 
    
\end{tabular}
\label{tab:nophantom_thetayscan_multiphi}
\end{table}

\subsection{Rotation over three axes}

\label{sec:without_phantom_rotXYZ}
A more complicated case was studied where, first, 
    the monitor sources are rotated with respect to the rings with angle $\Delta \Phi = -30$\degr.
Hence, the point sources are now at positions where their $\Phi$ angles are 
    330\degr, 90\degr \ and 210\degr \ for the first ring (instead of the usual 0\degr, 120\degr \ and 240\degr)
    and 30\degr, 150\degr \ and 270\degr \ for the second ring (instead of the usual 60\degr, 180\degr \ and 300\degr).
Next, the rings were rotated first around the y-axis with angle $\Theta(y) = 10$\degr, 
    next around the x-axis with angle $\Theta(x) = 20$\degr. 
Combining the two $\Theta$ rotations results in an overall polar angle of $\Theta = 22.3$\degr. 
The measured values for $\Theta$ are shown in table~\ref{tab:without_phantom_rotXYZ_thetaXYA}.
Table~\ref{tab:without_phantom_rotXYZ_phi} gives an overview of the measured values for \dPHI.
\begin{table}[tph!]
\caption{Mean estimated deviation of $\Theta$ and RMS averaged over 10 sets of 500 coincidences.}
\begin{tabular}{|l|l|l|l|}
\hline
                         & \textbf{$\Theta$(X)}      & \textbf{$\Theta$(Y)}  & \textbf{$\Theta$}  \\ \hline
\textbf{true}            & 20\degr            & 10\degr        & 22.3\degr        \\ \hline
\textbf{measured}        & 19.3$ \pm $0.1\degr   & 9.7$ \pm $0.1\degr  & 21.5$ \pm $0.1\degr    \\ \hline
\textbf{deviation}       &  0.7$ \pm $0.1\degr   & 0.3$ \pm $0.1\degr  &  0.8$ \pm $0.1\degr   \\ \hline
\textbf{arc length [mm]} &  1.2$ \pm $0.2        & 0.4$ \pm $0.2       &  1.3$ \pm $0.2    \\ \hline
\end{tabular}
\label{tab:without_phantom_rotXYZ_thetaXYA}
\end{table}
\begin{table}[tph!]
\caption{Mean estimated deviation of \dPHI and RMS averaged over 10 sets of 500 coincidences.}
\begin{tabular}{|l|l|l|l|l|l|l|}
\hline
                            & \multicolumn{6}{l|}{\textbf{True \dPHI} = 30\degr }  \\ \hline
\textbf{measured}           & 29.6$ \pm $0.2\degr  & 30.2$ \pm $0.1\degr   & 30.0$ \pm $0.2\degr 
                            & 29.6$ \pm $0.3\degr  & 30.1$ \pm $0.3\degr   & 29.9$ \pm $0.2\degr \\ \hline
                            
\textbf{deviation}          & -0.4$ \pm $0.2\degr  & 0.2$ \pm $0.1\degr   & 0.0$ \pm $0.2\degr          
                            & -0.5$ \pm $0.3\degr  & 0.1$ \pm $0.3\degr   & -0.1$ \pm $0.2\degr          \\ \hline
                            
\textbf{arc length [mm]}    & -0.7$ \pm $0.3   & 0.4$ \pm $0.2    & 0.1$ \pm $0.3   
                            & -0.7$ \pm $0.4   & 0.2$ \pm $0.4    & -0.2$ \pm $0.4   \\ \hline
\textbf{<arc length>}            
        & \multicolumn{6}{l|}{-0.2$ \pm $0.3 mm}  \\ \hline
\end{tabular}
\label{tab:without_phantom_rotXYZ_phi}
\end{table}

%
%

\subsection{Rotation over three axes, variation of crystal properties}

\label{sec:without_phantom_rotXYZ_dEE_size}

The next simulations were done also with $\Delta \Phi = -30$\degr, $\Theta(y) = 10$\degr, 
    and $\Theta(x) = 20$\degr \, but changing some properties of the scanner crystal.
%


\subsubsection{Variation of energy resolution}
\label{sec:without_phantom_rotXYZ_dEE}

First, instead of using an energy resolution of 10\% FWHM, 
    we assigned an alternative energy resolution of 15 \% FWHM to the crystals.
Table~\ref{tab:without_phantom_rotXYZ_thetaXYA_dEE} shows results for the estimated $\Theta$ values.
Table~\ref{tab:without_phantom_rotXYZ_phi_dEE} gives an overview of the measured values for \dPHI.    
\begin{table}[tph!]
\caption{Mean estimated deviation of $\Theta$ and RMS averaged over 10 sets of 500 coincidences 
    in the case of an alternative energy resolution of 15 \% FWHM.
    The table should be compared with table~\ref{tab:without_phantom_rotXYZ_thetaXYA}.
    }
\begin{tabular}{|l|l|l|l|}
\hline
                         & \textbf{$\Theta$(X)}      & \textbf{$\Theta$(Y)}  & \textbf{$\Theta$}  \\ \hline
\textbf{true}            & 20\degr           & 10\degr          & 22.3\degr         \\ \hline
\textbf{measured}        & 19.2$ \pm $0.1\degr   & 9.8$ \pm $0.1\degr   & 21.5$ \pm $0.1\degr \\ \hline
\textbf{deviation}       & 0.8$ \pm $0.1\degr    & 0.2$ \pm $0.1\degr   & 0.8$ \pm $0.1\degr         \\ \hline
\textbf{arc length [mm]} & 1.3$ \pm $0.1         & 0.4$ \pm $0.2   & 1.3$ \pm $0.2  \\ \hline
\end{tabular}
\label{tab:without_phantom_rotXYZ_thetaXYA_dEE}
\end{table}
\begin{table}[tph!]
\caption{Mean estimated deviation of \dPHI and RMS averaged over 10 sets of 500 coincidences
    in the case of an alternative energy resolution of 15 \% FWHM.
    The table should be compared with table~\ref{tab:without_phantom_rotXYZ_phi}.
    }
\begin{tabular}{|l|l|l|l|l|l|l|}
\hline
                         & \multicolumn{6}{l|}{\textbf{True \dPHI} = 30\degr }  \\ \hline
\textbf{measured}        & 29.5$ \pm $0.2\degr & 30.2$ \pm $0.1\degr & 29.9$ \pm $0.2\degr 
                         & 29.5$ \pm $0.2\degr & 30.2$ \pm $0.2\degr & 30.0$ \pm $0.2\degr \\ \hline
                         
\textbf{deviation}       & -0.5$ \pm $0.2\degr  & 0.2$ \pm $0.1\degr  & -0.1$ \pm $0.2\degr        
                         & -0.5$ \pm $0.2\degr  & 0.2$ \pm $0.2\degr  & 0.0$ \pm $0.2\degr         \\  \hline
                         
\textbf{arc length [mm]} & -0.8$ \pm $0.4       & 0.4$ \pm $0.2       & -0.2$ \pm $0.3       
                         & -0.8$ \pm $0.4       & 0.3$ \pm $0.3       & 0.0$ \pm $0.3  \\ \hline
\textbf{<arc length>}            
        & \multicolumn{6}{l|}{-0.2$ \pm $0.3 mm}  \\ \hline   
\end{tabular}
\label{tab:without_phantom_rotXYZ_phi_dEE}
\end{table}
%


\subsubsection{Variation of crystal size}
\label{sec:without_phantom_rotXYZ_size}

Next, we changed the size of the crystals to $3 \times 3 \times 20$ mm$^3$.
Table~\ref{tab:without_phantom_rotXYZ_thetaXYA_3x3} shows results for the estimated $\Theta$ values.
Table~\ref{tab:without_phantom_rotXYZ_phi_3x3} gives an overview of the measured values for \dPHI.    
\begin{table}[tph!]
\caption{Mean estimated deviation of $\Theta$ and RMS averaged over 10 sets of 500 coincidences 
    in the case of a crystal size of $3 \times 3 \times 20$ mm$^3$.
    The table should be compared with table~\ref{tab:without_phantom_rotXYZ_thetaXYA}.
    }
\begin{tabular}{|l|l|l|l|}
\hline
                         & \textbf{$\Theta$(X)}      & \textbf{$\Theta$(Y)}  & \textbf{$\Theta$}  \\ \hline
\textbf{true}            & 20\degr           & 10\degr          & 22.3\degr         \\ \hline
\textbf{measured}        & 19.2$ \pm $0.1\degr   & 9.7$ \pm $0.1\degr   & 21.5$ \pm $0.1\degr \\ \hline
\textbf{deviation}       & 0.8$ \pm $0.1\degr    & 0.3$ \pm $0.1\degr   & 0.8$ \pm $0.1\degr         \\ \hline
\textbf{arc length [mm]} & 1.3$ \pm $0.2         & 0.5$ \pm $0.2        & 1.4$ \pm $0.2  \\ \hline
\end{tabular}
\label{tab:without_phantom_rotXYZ_thetaXYA_3x3}
\end{table}
\begin{table}[tphb!]
\caption{Mean estimated deviation of \dPHI and RMS averaged over 10 sets of 500 coincidences
    in the case of a crystal size of $3 \times 3 \times 20$ mm$^3$.
    The table should be compared with table~\ref{tab:without_phantom_rotXYZ_phi}.
    }
\begin{tabular}{|l|l|l|l|l|l|l|}
\hline
                & \multicolumn{6}{l|}{\textbf{True \dPHI} = 30\degr }   \\ \hline
\textbf{measured}        & 29.5$ \pm $0.2\degr  & 30.1$ \pm $0.2\degr & 30.0$ \pm $0.2\degr  
                         & 30.1$ \pm $0.2\degr  & 30.2$ \pm $0.2\degr & 29.8$ \pm $0.2\degr \\ \hline
                         
\textbf{deviation}       & -0.5$ \pm $0.2\degr  & 0.1$ \pm $0.2\degr  &  0.0$ \pm $0.2\degr        
                         &  0.1$ \pm $0.2\degr  & 0.2$ \pm $0.2\degr  & -0.2$ \pm $0.2\degr        \\ \hline
                         
\textbf{arc length [mm]} & -0.8$ \pm $0.3      & 0.2$ \pm $0.3       &  0.0$ \pm $0.3      
                         &  0.7$ \pm $0.4      & 0.4$ \pm $0.4       & -0.3$ \pm $0.3  \\ \hline

\textbf{<arc length>}            
        & \multicolumn{6}{l|}{0.0$ \pm $0.3 mm}  \\ \hline   
                         
\end{tabular}
\label{tab:without_phantom_rotXYZ_phi_3x3}
\end{table}
%

%
%

\subsection{Rotation over three axes and axial shift}

\label{sec:without_phantom_rotXYZ_shiftZ}

First, an axial shift of both rings by $\Delta Z = 3$ mm was studied.
The axial shift is estimated by taking the average z position 
    between the centres of the two fitted planes: $\Delta Z = (z1 + z2)/2$.
%
%
The average over 10 sets of 500 coincidences gives $\Delta Z = 3.0 \pm 0.1$ mm 
    and hence an average deviation of $0.0 \pm 0.1$ mm.
Tables~\ref{tab:without_phantom_rotXYZ_thetaXYA_dZ3mm} and~\ref{tab:without_phantom_rotXYZ_phi_dZ3mm}
    show results for the estimated values of $\Theta$ and \dPHI respectively.
\begin{table}[tph!]
\caption{Mean estimated deviation of $\Theta$ and RMS averaged over 10 sets of 500 coincidences 
    in the case of an axial shift of the head by 3 mm.}
\begin{tabular}{|l|l|l|l|}
\hline
                         & \textbf{$\Theta$(X)}       & \textbf{$\Theta$(Y)}  & \textbf{$\Theta$}  \\ \hline
\textbf{target}          & 20\degr           & 10\degr          & 22.3\degr         \\ \hline
\textbf{measured}        & 19.2$ \pm $0.1\degr   & 9.7$ \pm $0.1\degr   & 21.5$ \pm $0.1\degr \\ \hline
\textbf{deviation}       & 0.8$ \pm $0.1\degr    & 0.3$ \pm $0.1\degr   & 0.9$ \pm $0.1\degr         \\ \hline
\textbf{arc length [mm]} & 1.3$ \pm $0.11        & 0.5$ \pm $0.2        & 1.4 + 0.2  \\ \hline
\end{tabular}
\label{tab:without_phantom_rotXYZ_thetaXYA_dZ3mm}
\end{table}
\begin{table}[tph!]
\caption{Mean estimated deviation of \dPHI and RMS averaged over 10 sets of 500 coincidences
    in the case of an axial shift $\Delta z = 3$ mm.}
\begin{tabular}{|l|l|l|l|l|l|l|}
\hline
                         & \multicolumn{6}{l|}{\textbf{True \dPHI} = 30\degr }  \\ \hline
\textbf{measured}        & 29.5$ \pm $0.2\degr  & 30.2$ \pm $0.1\degr  & 29.8$ \pm $0.2\degr  
                         & 29.7$ \pm $0.2\degr  & 30.3$ \pm $0.1\degr  & 29.9$ \pm $0.2\degr \\ \hline
                         
\textbf{deviation}       & -0.5$ \pm $0.2\degr  & 0.2$ \pm $0.1\degr   & -0.2$ \pm $0.2\degr        
                         & -0.3$ \pm $0.2\degr  & 0.3$ \pm $0.1\degr   & -0.1$ \pm $0.2\degr        \\ \hline
                         
\textbf{arc length [mm]} & -0.8$ \pm $0.3        & 0.4$ \pm $0.2        & -0.3$ \pm $0.3       
                         & -0.5$ \pm $0.3       & 0.5$ \pm $0.2        & -0.2$ \pm $0.4 \\ \hline

\textbf{<arc length>}            
        & \multicolumn{6}{l|}{-0.2$ \pm $0.3 mm}  \\ \hline   
        
\end{tabular}
\label{tab:without_phantom_rotXYZ_phi_dZ3mm}
\end{table}
%
%
Next, an axial shift of $\Delta Z = 6$ mm was studied.
The average over 10 sets of 500 coincidences gives $\Delta Z = 5.9 \pm 0.1$ mm and 
    hence a deviation of $0.1 \pm 0.1$ mm.
Tables~\ref{tab:without_phantom_rotXYZ_thetaXYA_dZ6mm} and~\ref{tab:without_phantom_rotXYZ_phi_dZ6mm}
    show results for the estimated values of $\Theta$ and \dPHI respectively.
\begin{table}[tph!]
\caption{Mean estimated deviation of $\Theta$ and RMS averaged over 10 sets of 500 coincidences 
    in the case of an axial shift of the head by 6 mm.}
\begin{tabular}{|l|l|l|l|}
\hline
                         & \textbf{$\Theta$(X)}       & \textbf{$\Theta$(Y)}  & \textbf{$\Theta$}  \\ \hline
\textbf{true}            & 20\degr           & 10\degr          & 22.3\degr           \\ \hline
\textbf{measured}        & 19.2$ \pm $0.1\degr   & 9.7$ \pm $0.1\degr   & 21.5$ \pm $0.1\degr   \\ \hline
\textbf{deviation}       & 0.8$ \pm $0.1\degr    & 0.3$ \pm $0.1\degr   & 0.8$ \pm $0.1\degr           \\ \hline
\textbf{arc length [mm]} & 1.3$ \pm $0.2         & 0.5$ \pm $0.2        & 1.4$ \pm $0.2    \\ \hline
\end{tabular}
\label{tab:without_phantom_rotXYZ_thetaXYA_dZ6mm}
\end{table}
\begin{table}[tph!]
\caption{Mean estimated deviation of \dPHI and RMS averaged over 10 sets of 500 coincidences
    in the case of an axial shift $\Delta z = 6$ mm.}
\begin{tabular}{|l|l|l|l|l|l|l|}
\hline
                         & \multicolumn{6}{l|}{\textbf{True \dPHI} = 30\degr }                                    \\ \hline
\textbf{measured}        & 29.7$ \pm $0.2\degr  & 30.1$ \pm $0.1\degr & 29.8$ \pm $0.1\degr  
                         & 29.7$ \pm $0.2\degr  & 30.1$ \pm $0.2\degr  & 30.0$ \pm $0.1\degr \\ \hline
                         
\textbf{deviation}       & -0.3$ \pm $0.2\degr  & 0.1$ \pm $0.1\degr  & -0.2$ \pm $0.1\degr        
                         & -0.3$ \pm $0.2\degr  & -0.1$ \pm $0.1\degr &  0.0$ \pm $0.1\degr      \\ \hline
                         
\textbf{arc length [mm]} & -0.5$ \pm $0.3       & 0.2$ \pm $0.2       & -0.3$ \pm $0.2       
                         & -0.5$ \pm $0.3       & -0.2$ \pm $0.3      &  0.0$ \pm $0.2  \\ \hline

\textbf{<arc length>}            
        & \multicolumn{6}{l|}{-0.2$ \pm $0.3 mm}  \\ \hline   
        
\end{tabular}
\label{tab:without_phantom_rotXYZ_phi_dZ6mm}
\end{table}
%
%
Finally, an axial shift of $\Delta Z = 10$ mm was studied.
The average over 10 sets of 500 coincidences gives $\Delta Z = 9.9 \pm 0.1$ mm 
    and hence a deviation of $0.1 \pm 0.1$ mm.
Tables~\ref{tab:without_phantom_rotXYZ_thetaXYA_dZ10mm} and~\ref{tab:without_phantom_rotXYZ_phi_dZ10mm}
    show results for the estimated values of $\Theta$  and \dPHI respectively.
\begin{table}[tph!]
\caption{Mean estimated deviation of $\Theta$ and RMS averaged over 10 sets of 500 coincidences 
    in the case of an axial shift of the head by 10 mm.}
\begin{tabular}{|l|l|l|l|}
\hline
                         & \textbf{$\Theta$(X)}       & \textbf{$\Theta$(Y)}  & \textbf{$\Theta$}  \\ \hline
\textbf{true}            & 20\degr           & 10\degr          & 22.3\degr         \\ \hline
\textbf{measured}        & 19.2$ \pm $0.1\degr   & 9.8$ \pm $0.1\degr   & 21.5$ \pm $0.1\degr \\ \hline
\textbf{deviation}       & 0.8$ \pm $0.1\degr    & 0.3$ \pm $0.1\degr   & 0.8$ \pm $0.1\degr         \\ \hline
\textbf{arc length [mm]} & 1.3$ \pm $0.2         & 0.4$ \pm $0.2        & 1.4$ \pm $0.2        \\ \hline
\end{tabular}
\label{tab:without_phantom_rotXYZ_thetaXYA_dZ10mm}
\end{table}
\begin{table}[tph!]
\caption{Mean estimated deviation of \dPHI and RMS averaged over 10 sets of 500 coincidences
    in the case of an axial shift $\Delta z = 10$ mm.}
\begin{tabular}{|l|l|l|l|l|l|l|}
\hline
                         & \multicolumn{6}{l|}{\textbf{True \dPHI} = 30\degr }                                    \\ \hline
\textbf{measured}        & 29.6$ \pm $0.2\degr  & 30.1$ \pm $0.2\degr  & 29.8$ \pm $0.2\degr  
                         & 29.5$ \pm $0.3\degr  & 30.0$ \pm $0.2\degr  & 30.0$ \pm $0.2\degr \\ \hline
                         
\textbf{deviation}       & -0.4$ \pm $0.2\degr  & 0.1$ \pm $0.2\degr   & -0.2$ \pm $0.2\degr        
                         & -0.5$ \pm $0.3\degr  & 0.0$ \pm $0.2\degr   &  0.0$ \pm $0.2\degr        \\ \hline
                         
\textbf{arc length [mm]} & -0.7$ \pm $0.3       & 0.2$ \pm $0.3        & -0.3$ \pm $0.3       
                         & -0.8$ \pm $0.5       & 0.0$ \pm $0.4        &  0.0$ \pm $0.3 \\ \hline

\textbf{<arc length>}            
        & \multicolumn{6}{l|}{-0.3$ \pm $0.4 mm}  \\ \hline   
        
\end{tabular}
\label{tab:without_phantom_rotXYZ_phi_dZ10mm}
\end{table}

%
%

\subsection{Alternative distribution of point sources over monitor rings}

\label{sec:without_different_PHIs}

\begin{figure}[tph!]
    \centering
    \begin{tabular}{l r}
    \includegraphics[width=0.36\linewidth]{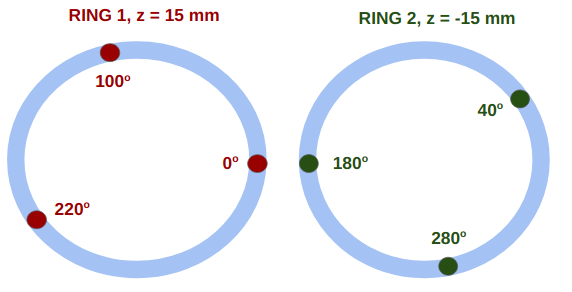}
    \end{tabular}
\caption{
Alternative distribution of point sources along the two rings. }
\label{fig:different_PHIs}
\end{figure}

%
%
%
Additionally, the point sources were put at different locations along their rings, 
    as shown in figure~\ref{fig:different_PHIs}.
Next, the entire system of rings was rotated 20\degr \ around the x-axis and 10\degr \ around the y-axis.
Without additional rotation in the x-y frame (the point sources have the assigned locations on the rings), we expect to find \dPHI = 0.
The results for the estimated angles $\Theta$ and \dPHI are shown in tables~\ref{tab:arghh_thetaXYA} and~\ref{tab:arghh_dPHI}.
\begin{table}[tph!]
\caption{Mean estimated deviation of $\Theta$ and RMS averaged over 10 sets of 500 coincidences 
    in the case of an alternative configuration of point sources on the rings.}
\begin{tabular}{|l|l|l|l|}
\hline
                         & \textbf{$\Theta$(X)}       & \textbf{$\Theta$(Y)}  & \textbf{$\Theta$}  \\ \hline
\textbf{true}            & 20\degr           & 10\degr          & 22.3\degr         \\ \hline
\textbf{measured}        & 19.2$ \pm $0.1\degr   & 9.6$ \pm $0.1\degr   & 21.4$ \pm $0.1\degr \\ \hline
\textbf{deviation}       &  0.8$ \pm $0.1\degr   & 0.4$ \pm $0.1\degr   & -0.8$ \pm $0.1\degr        \\ \hline
\textbf{arc length [mm]} &  1.4$ \pm $0.1        & 0.6$ \pm $0.2        &  1.4$ \pm $0.1  \\ \hline
\end{tabular}
\label{tab:arghh_thetaXYA}
\end{table}
\begin{table}[tph!]
\caption{Mean estimated deviation of \dPHI and RMS averaged over 10 sets of 500 coincidences
    in the case of an alternative configuration of point sources on the rings.
All angles are given in degrees.}
\begin{tabular}{|l|l|l|l|l|l|l|}
\hline
                         & \multicolumn{6}{l|}{\textbf{True \dPHI} = 0\degr }                                    \\ \hline
\textbf{measured}        & 0.5$ \pm $0.2\degr   & -0.1$ \pm $0.1\degr  & 0.1$ \pm $0.2\degr  
                         & 0.4$ \pm $0.3\degr   & -0.3$ \pm $0.3\degr  & 0.1$ \pm $0.1\degr \\ \hline
                         
\textbf{deviation}       & 0.5$ \pm $0.2\degr   & -0.1$ \pm $0.1\degr  & 0.1$ \pm $0.2\degr        
                         & 0.4$ \pm $0.3\degr   & -0.3$ \pm $0.3\degr  & 0.1$ \pm $0.1\degr \\ \hline
                         
\textbf{arc length [mm]} & 0.9$ \pm $0.3        & -0.2$ \pm $0.2       & 0.2$ \pm $0.3       
                         & 0.7$ \pm $0.4        & -0.4$ \pm $0.5       & 0.2$ \pm $0.2 \\ \hline

\textbf{<arc length>}            
        & \multicolumn{6}{l|}{0.2$ \pm $0.3 mm}  \\ \hline   
        
\end{tabular}
\label{tab:arghh_dPHI}
\end{table}

\section{Discussion}

In this analysis we have evaluated the performance of the CrowN@22 device 
	to track movements of a patient's head in a dedicated brain PET scanner. 
The evaluation is done by simulations on a configuration of two monitor ring, 
    where each ring has three $^{22}$Na point sources 
    and the rings are separated by 30 mm in the axial direction. 
Other non-pure positron emission isotopes such as $^{44}$Sc could be used  
	but $^{22}$Na is particularly suitable since it has a long half-life time (2.6 years) 
	and a small positron energy leading to a short mean free path before annihilation.
%

%
%
The results show that the CrowN@22 device can estimate a shift in z 
    and a rotation angle $\Phi$ in the x-y frame with excellent accuracy.
%
%
For our estimate of the $\Phi$ angle the error is less than 0.1 degree.   
The small deviations in the estimation of the $\Theta(y)$ angle in the x-z frame before correction 
    (table~\ref{tab:multitheta_multicoinc_results})
    show a linear behaviour which can be corrected for.
%
Table~\ref{tab:corrected_arc_lengths} shows that applying the correction  
    would allow the device to detect a movement 
    with a precision of less than 0.3 degrees or 0.5 mm in the $\Theta(y)$ direction.
%
We have also shown that for relatively low $^{22}$Na activities and small number of triple coincidences, 
    excellent estimations for the angles can be obtained.
These results show that the CrowN@22 technique is a reliable method 
    to monitor the motion of a patient's head inside a PET scanner.
%

%
%
{
For our results, we generated 20 M events with a total activity of 75.6 MBq ($^{18}$F plus $^{22}$Na)
    within a total event generation time of $0.265$ s, 
    where the total activity of the 6 $^{22}$Na point sources was 600 kBq. 
%
%
Table~\ref{tab:phantom_coinc_efficiencies} shows we find about 1500 triple coincidences, 
    which results in a counting rate (CR) of 5660 events per second.
%
%
Scaling down to 60 kBq total $^{22}$Na activity 
    would result in a counting rate of $\approx$ 570 events per second.
Section~\ref{sec:with_phantom} shows that less than 500 triple coincidences are needed
    for an accurate and precise measurement  
    so we conclude that the monitor measurements could be done with a 1 Hz sampling rate.
    }
%
%
%

%
%
The simulations presented in this article serve as a proof of concept.
Whereas the concept of CrowN@22 can be applied to most PET scanners, 
	the actual physical design and ergonomics can change according to realistic considerations
	without changing the conclusions presented here.
%
It would not be realistic to evaluate via simulation the performance of CrowN@22 
    on the wide range of possible brain PET scanners. 
Changing the energy resolution or crystal size showed no significant differences in the estimation 
    of the $\Theta$ and $\Phi$ angles.
%
%

%

%
The precise details of the CrowN@22 device, 
    such as the shape and material of the rings and whether they are exactly parallel, 
    will not affect the accuracy of the results.
Each ring with 3 point sources defines a unique plane which imposes a strong fiducial constraint 
    on its location in space and time.
This allows for precise and accurate tracking of the head movement 
    since the signal from the ring is well distinguishable from the background. 
Simulations with one single ring with 6 point sources gave similar results. 
The next step in our research will be the preparation of a CrowN@22 prototype, 
    using a Flexible Silicone Wig Headband which does not slip 
    and is made from elastic material in the form of a band, 
    to experimentally evaluate the CrowN@22 design.

\section{Conclusions}

We present a simulation analysis of the CrowN@22 device to monitor the head motion 
    in a scintillator brain PET scanner. 
The device is made of a set of $^{22}$Na point sources placed in two parallel rings
    separated by 30 mm in the axial direction. 
%
With very low $^{22}$Na activities of the order of 10 kBq per point source, 
      the precision of this device is less than 0.3 degrees or 0.5 mm at a 1 Hz sampling rate.
%
%
The CrowN@22 device serves as a co-registration marker for both CT and PET 
    to assist in carrying out the attenuation corrections accurately.

\section*{Acknowledgments}


%

This work acknowledges financial support 
    from the Spanish Ministry of Science and Innovation (MICINN) 
    through the Spanish State Research Agency, 
    under Severo Ochoa Centres of Excellence Programme 2025-2029 (CEX2024001442-S).

\bibliographystyle{JHEP}

\bibliography{8_biblio}

\end{document}